\documentstyle[12pt,epsf, rotate]{article}
\def\gsim{\mathrel{\rlap {\raise.5ex\hbox{$ > $}}
{\lower.5ex\hbox{$\sim$}}}}
\def\lsim{\mathrel{\rlap {\raise.5ex\hbox{$ < $}}
{\lower.5ex\hbox{$\sim$}}}}

\newcommand{\be}{\begin{equation}}
\newcommand{\ee}{\end{equation}}
\newcommand{\bea}{\begin{eqnarray}}
\newcommand{\nn}{\nonumber}
\newcommand{\eea}{\end{eqnarray}}

\def\gappeq{\mathrel{\rlap {\raise.5ex\hbox{$>$}}
{\lower.5ex\hbox{$\sim$}}}}
 
\def\lappeq{\mathrel{\rlap{\raise.5ex\hbox{$<$}}
{\lower.5ex\hbox{$\sim$}}}}
 
\begin{document}
 
\begin{titlepage}
\begin{flushright}

CERN-TH/99--43 \\
OUTP--99--14P \\
hep-th/9902190 \\
\end{flushright}

\begin{centering}
\vspace{.1in}

{\Large\bf{On the Thermodynamics of a Gas of AdS Black Holes
and the Quark-Hadron Phase Transition}}\\[15mm]

\vspace{.1in}

{\bf John Ellis}$^{a}$, 
{\bf A. Ghosh}$^{a}$ 
and {\bf N.E. Mavromatos}$^{a,b,\diamond}$ \\[.5cm]

$^{a}$ CERN, Theory Division, Geneva CH-1211, Geneva 23, Switzerland.\\
$^{b}$ University of Oxford, 
Department of Physics, Theoretical Physics,
1 Keble Road,
Oxford OX1 3NP, U.K.  

\vspace{.5in}
 
{\bf Abstract} \\
\vspace{.1in}
\end{centering}
{\small }
We discuss the thermodynamics of a gas of black holes in
five-dimensional anti-de-Sitter (AdS) space, showing that they are
described by a van der Waals equation of state. Motivated by
the Maldacena conjecture, we relate the energy density and pressure
of this non-ideal AdS black-hole gas to those of 
four-dimensional gauge theory in the unconfined phase. We 
find that the energy density rises rapidly above the deconfinement
transition temperature, whilst the pressure rises more slowly
towards its asymptotic high-temperature value, in qualitative
agreement with lattice simulations.

\vspace{3.in}

\begin{flushleft}
$^{\diamond}$ P.P.A.R.C. Advanced Fellow.\\
\end{flushleft}

\end{titlepage} 
\section{Introduction} 

The striking conjecture of Maldacena~\cite{malda} 
on the equivalence of large-$N_c$ superconformal quantum
gauge theories on $d$-dimensional Minkowski space $M_d$ - considered as
the
boundary of $(d+1)$-dimensional anti-de Sitter space AdS$_{d+1}$ - to 
classical supergravity in the AdS bulk, has opened a new dialogue
between students of non-perturbative gauge theories and string theorists.
Quantities in the strong-coupling limit of gauge theory may be
calculable using classical correlators in AdS (super)gravity, and/or
non-perturbative aspects of string theory may be related to correlators
in gauge theories~\cite{gkp,witt1}, in a holographic spirit~\cite{holo}.
In particular, the AdS approach 
was used in~\cite{witt} to relax the assumption of 
four-dimensional supersymmetry by
starting from a supersymmetric theory in 
six dimensions, which was one of the cases for which the 
conjecture was thought to be valid, 
and compactifying appropriately two of the dimensions. 
The resulting compactification led to a
high-temperature regime for the four-dimensional 
boundary theory, which had broken supersymmetry. In this
way, confinement at low temperatures and deconfinement at high
temperatures could be demonstrated. However, the gauge theory
was still conformal, and asymptotic freedom was therefore not
present. Nevertheless, this approach has motivated intriguing
estimates of glueball masses~\cite{balls}, the quark-antiquark
potential~\cite{potential}
and QCD vacuum condensates~\cite{vacuum} that agree surprisingly well with
lattice and other phenomenological estimates.

Two of us (J.E. and N.E.M.) have proposed~\cite{em} a generalization
of this holographic approach to the AdS$_{d+1}$/$M_d$
correspondence which is based on Liouville string theory~\cite{ddk,emn}, in
which conformal symmetry and supersymmetry need not be assumed,
provided world-sheet defects~\cite{defects} are
taken into account properly in the Liouville-dressed
theory. 
The Liouville field itself provides an extra bulk dimension,
and the AdS structure is induced by
the recoil of the world-sheet defect, when considered in
interaction with a closed-string loop~\cite{kanti}.
Within this approach, it was possible to demonstrate
the formation of a condensate of world-sheet defects at
low temperatures, which was related to the condensation of magnetic
monopoles in target space and induced confinement~\cite{em}. It was
also possible to demonstrate the logarithmic running of the gauge
coupling strength. Although this Liouville approach is somewhat
heuristic, it opens up a new way to discuss non-supersymmetric QCD
in the strong-coupling regime and at finite temperature.
In particular, the target-space quark-hadron deconfinement-confinement
transition may be viewed as a 
Berezinskii-Kosterlitz-Thouless phase transition of
world-sheet vortices~\cite{sathiap,defects}, which can also be related to
the phase transition of black holes in AdS~\cite{hp}.

In this paper we embark on a heuristic attempt to
use this approach to model aspects of
quark-hadron phase transition.
Lattice analyses~\cite{satz} indicate that the free
energy of pure QCD rises relatively rapidly above the
critical temperature to approach the 
asymptotic ideal-gas value as predicted in perturbative QCD.
On the other hand, the pressure is calculated~\cite{satz} to
rise much more slowly towards its asymptotic ideal-gas value,
and one possible interpretation is that massive 
effective degress of freedom are important close to the transition,
causing a larger departure from the ideal-gas picture for the
pressure than is the case for the free energy. Calculations close to the
phase transition necessarily require non-perturbative techniques,
such as the lattice, and our hope is that the $M_4$/AdS$_5$ correspondence
may also prove useful in this region.

Specifically, we model aspects of the gluon plasma using a
non-ideal gas of black holes in AdS$_5$, interacting via
forces of van der Waals type, and described by an
effective van der Waals equation of state that we derive
in this paper.

In order to set this approach up in the most reliable way,
and to relate it most closely to previous work~\cite{witt},
we first consider the high-temperature limit. Here the
black holes in AdS$_5$ are stable as well as massive, and hence
suitable for interpretation as `molecules' of a non-relativistic gas
with small velocities $|u_i| \ll 1$. We demonstrate in this limit how
the AdS structure of the ambient bulk space-time itself may be
translated into non-ideal-gas interactions between the massive
black-hole `gas molecules'. The question then arises how to extend
this description down to lower temperatures.

According to the world-sheet point of view, target-space black holes
may be viewed as `spike' defects on the world sheet~\cite{defects}, which
are dual
to `vortex' defects, that in turn correspond to $D$ particles in
space-time~\cite{polch}. In the above-mentioned high-temperature limit,
these
$D$ particles are light, and difficult to treat using the
Liouville approach. However, the Liouville approach is well-adapted
for a discussion of a dual limit, in which the
$D$ particles become very heavy~\cite{em}.

It should be emphasized, though, that both of these limits correspond
to temperatures that are high compared to those in the confining phase.
In both QCD~\cite{CEO} and AdS gravity~\cite{hp}, three distinct
transition temperatures
have been identified: $T_0$, below which only the 
confined phase of the gauge theory exists
and black holes condense leaving a residual gas of radiation; $T_1$, at
which the free energies of the
confined and deconfined phases (or black-hole and radiation phases) are
equal; and $T_2$, beyond which only the
deconfined and stable-black-hole phases exist. In the world-sheet picture,
these correspond to the temperatures of Berezinskii-Kosterlitz-Thouless
transitions for vortex and spike condensation~\cite{em}. The
high-temperature
limit~\cite{witt} we take corresponds to $T \sim T_2$ and above, and our
lower-temperature limit corresponds to $T \gsim T_0$. We aim to
establish a judicious interpolation between these two limits
that describes qualitatively correctly the intermediate region $T \sim
T_1$.

The layout of this paper is as follows. In section 2 we discuss in
more detail general features of our approach to the thermodynamics
of AdS black holes. Then, in section 3 we discuss the high-energy limit
in which the $D$ particles are light~\cite{witt}, and section 4 contains
an application of Liouville string theory~\cite{em} to the
lower-temperature limit. Finally, in sections 5 and 6 we pull together a
general picture of the quark-hadron phase transition using the
information we have obtained from our studies of the thermodynamics
of AdS black holes, and relate it to lattice results~\cite{satz}.

\section{Thermodynamics of a Gas of AdS Black Holes}

We consider a homogeneous gas of AdS black holes,
each of mass $M$, and restrict ourselves to the case where the 
characteristic velocity of a generic black hole in the ensemble is 
either zero or very small: $|u_i|\ll 1$, corresponding to the 
case of very massive black holes. We consider first the
static case: $u_i=0$. This will help in gaining insight into the 
$u_i\ne 0$ case, which we treat later as a perturbation of
the static case. 

We consider an ensemble of $N$ indistinguishable 
Schwarzschild black holes in a five-dimensional AdS space-time
with radius $b$, which is related to the critical temperature $T_1$
above which massive black holes are stable: $T_1=1/(\pi b)$.
The invariant line element is be taken to be (in Minkowskian 
signature)~\cite{hp,witt}:
\be
ds^2=-\left(1 + \frac{r^2}{b^2} - \frac{\omega _4 M}{r^2}\right)dt^2 
+ \left(1 + \frac{r^2}{b^2} - \frac{\omega _4 M}{r^2}\right)^{-1}dr^2 
+ r^2 d\Omega_3 
\label{adsbh}
\ee
where the AdS radius $b$ is related to the negative cosmological
constant $\Lambda$ by $b=\sqrt{-3/\Lambda}$, and
$\omega_4 \equiv 8G_N/3\pi$, where $G_N$ is the five-dimensional 
Newton constant that is related to the Planck length $\ell_P$ via 
$G_N=\ell_P^3$, and $M$ is the ADM mass of the black hole. 
The outer horizon of the black hole is defined to be the 
larger positive root $r_+$ of the equation
\be
1 + \frac{r_+^2}{b^2}-\frac{\omega_4 M}{r_+^2}=0\,,
\label{rootequation}
\ee
namely
\be
r_+ =b\left( -\frac{1}{2} + \frac{1}{2}\sqrt{1 + 4\omega_4 M/b^2}\right)^{1/2}.
\label{rplus}\ee
For the purposes of calculating the partition function and other
thermodynamic quantities of the ensemble, a 
Wick rotation to a Euclidean AdS geometry:
$t \rightarrow i\,t$, will always be understood. 
The Euclideanized
AdS-Schwarzschild  space-time has been found to be 
smooth~\cite{hp}, provided the Euclidean time direction is perodic 
at a particular radius $\beta_0$:
\be
  \beta_0 \equiv   T_H^{-1}=\frac{4\pi b^2 r_+}{4 r_+^2 + 2b^2}\;.
\label{temp}
\ee
where $T_H$ is the Hawking temperature of the black hole.

In subsequent sections of this paper,
we consider {\it two} limits of (\ref{temp}) that
correspond to high Hawking temperatures $T_H \gsim T_0$, namely
(i) $b^2\ll r_+^2$ and (ii) $b^2\gg r_+^2$,
where we assume in each case that $\ell_P^2\ll r_+^2, b^2$.
It is easy to check using (\ref{rplus}) that in the limit (i) we also have
$\omega_4M/r_+^2 \sim r_+^2/b^2$, whereas in the limit (ii) we also have
$r_+^2\simeq\omega_4 M \gg \ell_P^2$, so that we are consistent with the
static limit in both cases.

According to the analysis of~\cite{hp},
the thermodynamic ensemble of black holes is stable
in the limit (i): $\ell_P^2\ll b^2\ll r_+^2\ll\omega_4 M$, 
which was considered in~\cite{witt}. This corresponds to 
the region $T \sim T_2$.
The limit (ii): $\ell_P^2\ll r_+^2\sim\omega_4 M\ll b^2$,
where the radius of the 
AdS space-time is large compared to the outer horizon,
was studied perturbatively in~\cite{em}, using the Liouville
approach where $b
\sim \delta^{-2} \rightarrow \infty$, with $\delta \rightarrow 0^+$ a
parameter appropriately defined to regulate
the recoil operators. According to~\cite{hp}, this limit corresponds to a
temperature $T \gsim T_0$. However,
the intermediate
regime of the phase transition where $T \sim T_1$, lying between the
regions (i) and (ii), 
cannot be studied reliably by analytic methods, and we resort later
to continuity arguments in order to describe the energy and pressure
curves in this region.

\section{Is there a Phase Transition?} 

To answer this question and to investigate its order,
one should examine the equation of state 
for an ensemble of AdS black holes in appropriate
regimes of the parameters. We consider first the limit (i) above.

Using standard General Relativity, the effective static 
potential $U(r)$ between two black holes in the ensemble
with a radial separation $r$ is given by the 
temporal metric component $g_{00}$, that can be obtained
from (\ref{adsbh}):  
\be
 g_{00}=-1-2U(r)\,, \qquad U(r)=\frac{r^2}{2b^2}  - 
\frac{\omega_4 M}{2r^2}
\label{grpotential}
\ee
provided the potential is weak. We note that the potential 
(\ref{grpotential}) vanishes at a radius $r_0$:
\be
U(r_0)=0\,\, \rightarrow \,\, r_0=\left(\omega_4 Mb^2\right)^{1/4}
\;.\ee
Using (\ref{rplus}), we see that $r_0\simeq\;r_++{b^2\over 4r_+}$\,.
Therefore, if we restrict ourselves to a thin shell outside the horizon:
$r_+\le r\le r_++\varepsilon$\,, the potential is indeed small, and
one can take $\varepsilon=(r_0-r_+)\simeq b^2/4r_+$ to
a good approximation.
Since the potential varies very 
little over the range $r_+\le r\le r_++\varepsilon$,
we make the second approximation $U(r)\simeq\overline
U$\,, where $\overline U$ is a constant, justified because 
$\delta U\sim\varepsilon$\,. The fact that the absolute value of
the constant $\overline U$ is a small number also justifies our 
weak-field approximation. These statements can be checked
more precisely using the following formulae: 
\bea 
&&U(r)\simeq{\overline U}=const\equiv\kappa N/\Delta\Omega\nn\\
&&\kappa=(1/N)\int_{r_+}^{r_++\varepsilon}d^4x\,U(r) 
\simeq{\pi^2\over 4N}\left[{\varepsilon r_+^5\over b^2}-\omega_4M\varepsilon r_+
\right]
\label{potent}
\eea
where the effective volume $\Delta\Omega=\Omega_4-\Omega_0$ is the size of 
the four-volume of the shell $r_+\le r\le r_++\varepsilon$ and $N$
is the number 
of black holes in the ensemble. We denote by $\Omega_0$ the
four-volume inside the horizon. From the expression 
(\ref{potent}) for $\kappa$, one can easily check that 
$\overline U<1$\,.
We shall see that the above approximations lead to
analytic expressions for thermodynamic quantities of our AdS 
black-hole system.

We first evaluate the classical partition function $Z$  
of the system, assuming that it is in thermal equilibrium 
at a Hawking temperature $T\equiv\beta^{-1}$:
\bea 
Z &=& \frac{1}{N!} 
\left(\int \frac{d^4 x d^4 p}{(2\pi)^4}\;e^{-\beta\big[p^2/2M + MU(r)
\big]}
\right)^N \nn \\
&=& \frac{1}{N!} \left(\frac{M}{2\pi\beta}\right)^{2N}\!\!\!\left(
\int d^4x\; e^{-\beta U(r) M }\right)^N\!\! = 
\frac{1}{N!}\left(\frac{M}{2\pi\beta}\right)^{2N}\!\!\Delta\Omega^N 
e^{-\beta {\overline U}MN }\;.
\label{part}
\eea
Some remarks are in order at this point. First, in the static
case which we are considering now, the kinetic term of the black
holes has no physical meaning and serves only as a cut-off for the 
momentum integrals. Secondly, the spatial integral is understood to
be taken over the prescribed shell $r_+\le r\le r_++\varepsilon$,
which give rise
to the volume factor $\Delta\Omega$. Finally, we underline that we have
made explicit use of the approximations mentioned above. 

Before proceeding to compute other thermodynamic quantities,
we also examine the low-velocity non-static case.
As we shall see, the difference between the static and the non static
cases can be described effectively by a renormalisation
shift of the mass term $M\to M+T$ in the partition function 
above (\ref{part}). 
To prove this, we note that in the non-static case, where the heavy 
black holes 
move with a small velocity $|u_i|\ll 1$ relative to each other, 
we may employ both the small-velocity and weak-potential
approximations simultaneously. To obtain the velocity corrections to
the potentials, we use the Minkowski-signature Lagrangian:
\be
L=-M\frac{ds}{dt}\,, \qquad ds^2=-g_{AB} dx^A dx^B,\qquad A,B=0...4
\label{timedep}
\ee
where $ds^2$ is given by (\ref{adsbh}). 
We then write
\be
 \frac{ds}{dt} = \sqrt{-g_{00} - g_{ij} \frac{d x^i}{dt} \frac{d x^j}{dt}}
=\sqrt{-g_{00} - g_{ij}u^i u^j}\;.
\label{form1}
\ee
and set
\be
    g_{00} =-1-2U(r)\,, \qquad g_{ij}=\delta _{ij} + h_{ij}\,,
\qquad r^2 \equiv \sum_{i=1}^4 x_i^2\;. 
\ee
{}from which we see that $ds/dt = \sqrt{1 + 2 U - |u_i|^2 - h_{ij} u^i u^j}$.
Expanding in powers of {\it both} U and $u_i^2$ to leading non-trivial
order, we find:
\be
    L \simeq -M \left(1 +  U -\frac{1}{2}|u_i|^2 
- \frac{1}{2} u^i h_{ij} u^j + 
\frac{1}{2}U |u_i|^2 +... \right)\;.
\label{Lexp}
\ee
Parametrising with a generalized velocity-dependent potential ${\tilde U}$ 
in the Lagrangian: $L=-M +\frac{1}{2}M|u_i|^2 - M {\tilde U}({\overline x}, u)$,
we get
\be
   {\tilde U} ({\overline x}, u) = U + \frac{1}{2}U|u_i|^2 
- \frac{1}{2} u^i h_{ij} u^j 
\label{genpot}
\ee
which clearly reduces to $U$
in the static limit $u_i \rightarrow 0$.

The partition function for slowly-moving black holes
can be computed in a straightforward manner. 
First, we note that the generalized momenta are given by
\be
 p_i = \frac{\partial L}{\partial u^i} = M
\left(u_i-u_iU+h_{ij}u^j\right),
\label{momenta}
\ee
giving rise to the Hamiltonian:
\be
   H = p_i u^i - L = M (1 + U) + \frac{1}{2}p_iu^i\;. 
\label{ham}
\ee
Taking into account the facts that $h_{ij}=2 Ux_i x_j/r^2$ 
and $u_i\simeq\frac{p_j}{M} [\delta _{ij}(1 + U) - h_{ij} ]$,
we can re-express the Hamiltonian as:
\be
  H= \frac{1}{2} p_i {\cal K}_{ij} p_j + MU\,, 
\qquad {\cal K}_{ij}=\frac{1}{M}\left[\delta _{ij} (1 + U) - h_{ij}\right]
 \label{ham2}
\ee
which resembles the Hamiltonian of a particle with momenta $p_i$ 
in a `curved space' whose `metric' ${\cal K}_{ij}$  
depends solely on the potential $U$ and not
on the velocity $u_i$ (to this order). 
The resulting partition function is 
\bea
Z&=&\frac{1}{N!} 
\left(\int \frac{d^4 x d^4 p}{(2\pi)^4}\, 
e^{-\beta\left[\frac{1}{2}p_i {\cal K}_{ij}p_j + MU(r)\right]}
\right)^N\nn\\ &=&
\frac{1}{N!}\left(\frac{M}{2\pi\beta}\right)^{2N}\!\!\left(
\int d^4x \frac{1}{\sqrt{{\rm det}(K M)}}
e^{-\beta U(r) M }\right)^N 
\label{part2}
\eea
where 
\bea
\sqrt{{\rm det}(KM)} = \left[{\rm Det}\big((1+U)\delta_{ij}-h_{ij}\big)
\right]^{1/2}\simeq\exp\left[\frac{1}{2}{\rm Tr}(U\delta_{ij}-h_{ij})+...
\right]\simeq e^U\;. 
\label{apprx2}
\eea
Thus, one has, upon approximating $U \simeq {\overline U}$ (c.f. 
(\ref{potent})):
\be
Z \simeq \frac{1}{N!}\left(\frac{M}{2\pi\beta}\right)^{2N}\!\!
\Delta\Omega^Ne^{-N (\beta M + 1){\overline U}}
\label{part3}
\ee
which, when compared with (\ref{part}), demonstrates the aforementioned
renormalisation shift in the mass by $T$\,. In view of this simple change,
{}from now on we shall deal with the general velocity-dependent case.

The energy of the system is defined as 
\be 
{\overline E} 
= \frac{\partial}{\partial \beta}\left(-{\rm Ln}Z + \beta \mu N \right)
\label{energyint}
\ee
where the `ground-state energy' due to the chemical potential 
$\mu=\partial{\rm Ln}Z/\partial N$ has been appropriately added.
The pressure of the system is defined as:
\be
     P \equiv \frac{1}{\beta} \frac{\partial {\rm Ln}Z}{\partial
\Delta \Omega},
\label{pressure}
\ee
which upon using (8) yields the following equation of state:
\be
NT\,=\,
\left(P-\kappa\tilde
M{N^2\over(\Omega_4-\Omega_0)^2}\right)\left(\Omega_4
-\Omega_0\right),
\label{eqstate}
\ee
where the Boltzmann constant has been put to unity and
$\tilde M$ represents the shifted mass
appearing in the partition function (\ref{part3}). 

The relation (\ref{eqstate}) is
nothing other than a {\it van der Waals equation of state}. In our view,
this leads to the {\it prima facie} expectation
that a first-order phase transition takes place in the bulk,
though this remains to be verified. 

For the non-static case (\ref{part3}),  
the quantity ${\overline E}$ is
\bea 
{\overline E}&=&-N{\rm Ln}N - N 
- 2N \frac{2\pi \beta}{M}\left(\frac{\partial M}{\partial \beta}\frac{1}
{2\pi \beta} - \frac{M}{2\pi \beta ^2}\right)(1 - \beta) 
+ 2 N {\rm Ln}(\frac{M}{2\pi\beta}) + \nn \\
&&N {\rm Ln}\Delta \Omega + \frac{N}{\Delta \Omega} \frac{\partial \Delta \Omega}{\partial \beta} 
+ ( 1 - 2\beta)MN{\overline U} - N {\overline U} \;.
\label{energyform}
\eea
In the limit under consideration, we may use (\ref{temp}) 
to relate the black-hole mass to the temperature:
\be
      M \sim \frac{\pi^4}{\omega_4}b^6 T^4
\label{mass}
\ee
for a five-dimensional AdS-Schwarzschild black hole.~\footnote{In
the case of an $(n+1)$-dimensional
AdS-Schwarzschild black hole of large mass $M$, the temperature scales
as $T^n$.}
We assume (see later) that the AdS radius $b$ scales with the temperature
as
\be
b\sim c_0T^{-1},
\label{c0}
\ee
where $c_0$ is taken sufficiently large to ensure 
that $\omega_4M\gg b^2$\,. Thus ${\overline E}$
is easily evaluated:
\be
{\overline E} = {\rm const} + 2 N T - 6N{\rm Ln}T 
+ c''{\overline U} N T^{-2} - 2 c''N {\overline U} T^{-3}  
\label{energy} 
\ee
where ${\overline U} < 0$, and is assumed constant, and we have used
${\rm Ln} N! \simeq N {\rm Ln} N$ for large $N$. 
The constant in (\ref{energy}) can be set to zero by an appropriate
normalization of the energy, since only energy differences matter.

The energy density $\rho$ for the {\it four-dimensional} system
on the boundary of AdS$_5$
is obtained by dividing ${\overline E}$ by the 
{\it three}-volume which, in view of the above discussion, scales 
like $T^{-3}$. Thus
\be 
\rho/T^4 \propto  2 N  - 6NT^{-1} {\rm Ln}T 
+ c''{\overline U} N T^{-3} - 2 c''N {\overline U} T^{-4} 
\label{energydens} 
\ee
As for the energy density, the pressure in
{\it four dimensions} (three space dimensions, one 
periodic temperature dimension), denoted by $p$,
is computed from the 
equation of state (\ref{eqstate}) after writing it 
in the form 
\be
p \equiv b P = {\rm const} \times \frac{1}{\Delta \Omega _3 
} NT ,
\label{fourdpre}
\ee
using (\ref{potent}) and assuming that the variations 
of the potential with the volume are small. 
The quantity $b P$ simply represents the fact that the 
three-space pressure should be computed with respect to a 
three-volume shell, $\Delta \Omega _3$, and {\it not} a 
four-volume shell, $\Delta \Omega _4$.
The former scales with one length dimension less compared to 
the latter, and thus the quantity $bp$ in (\ref{eqstate}) 
has the right scaling with $T$. 
With in this mind, we remark that $\Delta \Omega _3$ scales like $T^{-3}$
in the very-high-temperature regime, beyond the upper phase
transition, and hence that
\be
    p/T^4 \sim {\rm const} \;.
\label{pressure2} 
\ee
The constant term in (\ref{pressure2}) may be
fixed by the fact that in the very-high-temperature regime 
the system is supposed to represent a gas of massless gluons,
and hence, from the classical statistical mechanics of a ideal gas
of massless Bose particles,
the energy density is three times the pressure.

\begin{figure}[htb]
\epsfxsize=3in
\bigskip
\centerline{\epsffile{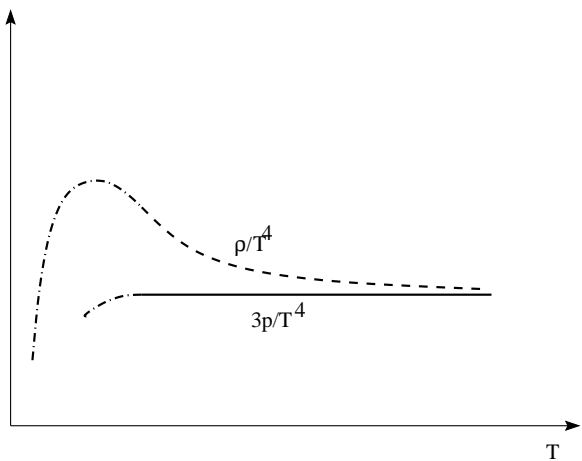}}
\caption{\it\baselineskip=12pt 
The scaled energy density $\rho/T^4$
(dashed line) and pressure $3p/T^4$
(continuous line) in a gauge theory, plotted as functions of the
temperature $T$, as
calculated in the high-temperature limit (i) $b \ll r_+$~\cite{witt}
using a typical set of parameters for $N$ indistinguishable AdS black
holes. The bump in the energy density is reminiscent of the 
transition from a gas of pions to a deconfined 
quark-gluon plasma in the QCD case, but the
approximations made in the limit (i) are not
reliable in the regions where the lines are dotted.}
\bigskip
\label{enerpress}\end{figure}

The energy density curve  
is plotted in Figure~\ref{enerpress}. 
We observe that the qualitative features of QCD
are correctly captured by our classical 
statistical system of AdS black holes.
The energy density drops sharply as we approach 
low temperatures, and it is tempting to identify 
this region might with 
the deconfined region of QCD, 
approached from the high-temperature unconfined
phase. Our approximate calculation exhibits
a bump in the energy-density curve
before the `confined' region is reached, due to the 
$-T^{-1} {\rm Ln }T$ term in (\ref{energydens}). 
However, the limit (i) that we have used above 
is valid only for high temperatures, and should not be trusted
quantitatively in this region. 
On the other hand, the appearance of this bump may 
indicate the existence of a thermodynamic instability, given that the 
`bump' region is followed by a sharp drop in the energy density as
the `confined' region is approached.

\section{Another View of the Phase Transition} 

In this section we shall look at the phase 
transition from the opposite viewpoint,
described by the limit (ii) defined above: $\ell_P^2\ll
r_+^2\simeq\omega_4 M\ll b^2$\,.
The approximations made in the analysis of the previous
section are also valid in this parameter regime,
though in a different way.
When there is a large separation between any pair of black holes in the
ensemble: $r_+\ll r\ll b$, it is again 
a good approximation to take $U(r)\simeq\overline U$\,,
where $\overline U$ is a positive constant, because the potential varies very 
little over the range $r_+\ll r\ll b$. Not only that, but the
constant $\overline U$ is also a small number and hence one can again make a
weak-field approximation. The analogues of the formulae (\ref{potent})
are in this case: 
\bea 
&&U(r)\simeq{\overline U}=const\equiv\kappa N/\Omega\nn\\
&&\kappa=(1/N)\int_{r_+}^bd^4x\,U(r) 
={\pi^2\over 24N}\left[b^4-{r_+^6\over b^2}-3\omega_4M(b^2-r_+^2)\right]
\label{potent2}
\eea
where, as before, the effective volume 
$\Omega=\Omega_4-\Omega_0$ is the size of 
the four-volume of the shell $r_+\ll r\ll b$, and $N$ is the
number of black holes in the ensemble. 
Given that $\Omega\sim b^4$, and using the above formula (\ref{potent2})
for $\kappa$, one can easily check that $\overline U<1$\,.
Notice that the potential is {\em attractive} in the
region $r_+\ll r<r_0$ and {\it repulsive} in the region $r_0<r\ll b$\,.
In this region, (\ref{temp}) tells us that the black-hole mass
is related to the temperature by:
\be
      M \sim \frac{\beta^2}{4\pi\omega_4}\;.
\label{mass2}
\ee
One can perform calculations for the partition function, energy
and pressure of the ensemble that are similar to those
described in the previous section, which we do not reproduce in detail
here.

As in the limit (i), we find again in the limit (ii)
a {\it van der Waals equation of state}, as in (\ref{eqstate}).

To understand qualitatively the physics in the
lower end of the transition
region $T \sim T_1$, we recall that, as we approach the
transition region from above,
we enter a regime where the Liouville theory takes over.
In this theory the radius $b$ may be assumed to be independent of
temperature~\cite{em}, and large. In this limit of
large $b$ and $T$-independence,
a different approximation is needed to capture correctly the
features of the transition region, since
the classical description of a gas
of stable black-hole particles breaks down in this case.
However, we can still obtain qualitative
ideas of the dynamics by applying the above
statistical-mechanical approach to this case. 
Notice that the smallness of $\omega_4 M $ compared to $b$
implies that the AdS space is regular for large $r$. This is the
regime discussed in the Liouville approach of~\cite{em}.

\begin{figure}[htb]
\epsfxsize=3in
\bigskip
\centerline{\epsffile{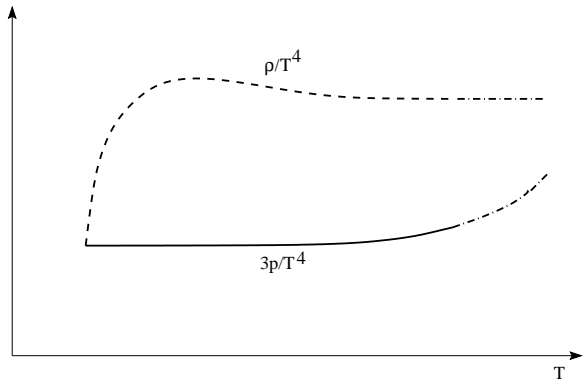}}
\caption{\it\baselineskip=12pt
As in Fig.~1, but
in the limit (ii): $r^+\ll b$~\cite{em}, conjectured to
represent the start of the phase transition regime.
The pressure curve lies below the energy density curve
and is almost constant: the bump in the energy density
is less marked than indicated in the limit (i).}
\bigskip
\label{adst2}\end{figure}

We have calculated
the energy density $\rho$ and the pressure $3p$,
with the results shown in Figure \ref{adst2}.
We observe that the pressure is almost constant near the transition region,
whilst the energy increases and exhibits a bump. As compared to our
results in limit (i), this bump is rather smoother.
The constant value of the pressure is again
fixed by the fact that at low temperatures the
system again enters an ideal-gas regime, where in this case
the physical degrees of freedom
are the bound states, i.e., massless pions in the case of QCD, so 
that the relation $\rho = 3p$ should again be valid.

\section{Relating the Two Descriptions}

The regime (i) which describes the high-temperature tail of the phase
transition
must connect smoothly with the regime (ii) as the scales cross:
$r_+\leftrightarrow
b$\,. Since one expects that the temperature should
rise after the phase transition,
we assume that $T_{(ii)}\ll T_{(i)}$\,. At the boundary of the two
regions, one has
$r_+^B\sim\sqrt{\omega_4 M_B}\sim b_B$, and the temperature $T_B\sim
1/b_B$ should therefore lie in the range
$T_{(ii)}\ll T_B\ll T_{(i)}$\,. A natural way to arrange this
crossover is to keep $r_+$ fixed at the boundary value and study the
variations of the other
parameters as we go from one regime to the other. Clearly 
this puts the following bounds on the other parameters:
\be
b_{(i)}\ll b_B\ll b_{(ii)}\,,\qquad M_{(ii)}\sim M_B\ll M_{(i)}\;.
\label{ordering}
\ee
These bounds seem natural and consistent with the definitions
of the limits (i) and (ii). In particular, one outcome of
this crossover, namely that the black-hole degrees of freedom
are more massive in 
region (i) and hence decouple from infrared
physics, seems consistent for describing the 
region just above the phase transition,
which, according to (\ref{ordering}), is associated with lower-mass black
holes: $M_{(ii)} \sim M_B \ll M_{(i)}$ in region (ii). 

We would also like to comment on the behaviour of the AdS radius $b$
in the transition region. 
The scaling (\ref{c0}) 
is justified in the Liouville approach of \cite{em}, 
in which the recoil-induced AdS radius $b$ is 
proportional to a homotopic `time' variable.
In the analysis of~\cite{em}, this homotopic time was identified
with the 
target time $X^0$ in a {\it real-time} formulation of Liouville QCD. 
In this real-time formalism, the time $X^0$ should not be confused
with the temperature. However, from the equivalence of the real-time and 
Matsubara formalisms, where one identifies the temperature with the 
inverse radius of a compactified Euclidean time, it is 
natural to assume that, at least in the high-temperature regime
where one assumes thermal equilibrium with a heat bath of temperature $T$, 
the scaling (\ref{c0}) should be valid. 

An alternative way to justify the scaling (\ref{c0}) is to
notice that, in order to arrive at the regime 
where the analysis of~\cite{em} 
is valid, one needs to go to very low temperatures, where $b$ is huge.
This result is not in contradiction 
with our above procedure of
identifying $b$ with $1/T$. However, in the low-temperature regime
$b$ is almost constant~\cite{em}, and {\it not} scaling with temperature.
We conjecture that there are
in general competing contributions to $b$, so that
\be
b \sim \delta ^{-2} + {\cal O}(1/T), 
\label{interpola}
\ee
and that the $\delta ^{-2}$ term dominates in the low-temperature
regime,
whilst $\delta $ is comparatively large in the high-temperature regime,
and the $1/T$ term dominates. 
In~\cite{em}, $\delta $ was identified with the 
area of the Wilson loop that generated the world sheet of the string.  
This is consistent with the above picture: for low temperatures
in the confining regime, the dominant 
degrees of freedom are related to large
world-sheet areas, in the sense that the 
(temporal Polyakov or spatial Wilson) loops that define the
order parameters relevant for confinement are large.
It is these quark-antiquark loops that can be described by 
weakly-coupled string theory, for which the analysis 
of~\cite{em} is valid. At
high temperatures, on the other hand, the areas defined by
the dominant order parameters (Wilson and/or Polyakov loops)
are relatively small or
microscopic, as remarked 
in~\cite{em}. This corresponds to the 
pure stringy limit $\delta ^{-2} \rightarrow 0$.
In that limit the perturbative string theory approach of~\cite{em} is
invalid, and should be replaced by the 
above semi-classical picture~\cite{witt} of a gas of very massive 
black holes. We now remark that, 
in our approximate treatment of near-horizon distances $r$, where the
potential is {\it weak}, one obtains the typical order of magnitude
estimate
\be
     r^4 \simeq \omega_4 M b^2 + {\cal O}(b^4\sqrt{\omega _4 M b^{-2}})
\label{untitle}
\ee
Combining (\ref{rplus}),(\ref{mass}) and (\ref{c0}), we find that
the volume $\Omega _0 \propto r_+^4$ varies with $T$ 
as $T^{-4}$, and hence 
\be
   \Delta \Omega \equiv \Omega_4 - \Omega _0 \propto r^4 - r_+^4 \simeq c' T^{-4} 
\label{domega}
\ee
where $c'$ is a constant.

As commented above and in~\cite{hp}, the phase transition 
of the five-dimensional black-hole system is 
expected to be of first order. Moreover, it is
here identified with a deconfining transition in gauge theory. 
However, at present our analysis cannot determine the precise order of
these associated transitions, and this remains an open issue. 
A related issue is
whether holography~\cite{holo} survives the first-order phase transitions
associated with the boundary and bulk dynamics. Based on the Liouville
renormalization argument given above, we would expect so, but this
issue is also open.

\section{Comparison with QCD} 

\begin{figure}[htb]
\epsfxsize=3in
\bigskip
\centerline{\epsffile{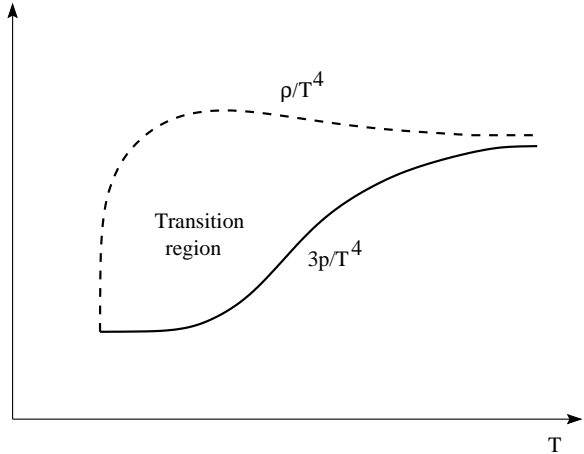}}
\caption{\it\baselineskip=12pt 
Interpolations of the scaled
energy density $\rho/T^4$ (dashed line) 
and pressure $3p/T^4$ 
(continuous line),
including the transition region between the limits (i) and (ii)
shown in Figs.~1 and 2, respectively.
The behaviours of the energy density and pressure
in the intermediate-temperature region
are reminiscent of lattice calculations~\cite{satz}:
in particular, the pressure curve rises more slowly than that of the
energy density.}
\bigskip
\label{threepre}\end{figure}

Here we comment on the temperature-dependence of the pressure, and relate
it to what is known for QCD from lattice simulations.
We recall from the discussion of section 4 that  
in low-temperature limit (i) of the phase 
transition region (see Fig.~\ref{threepre}), 
where $b$ is roughly $T$-independent and 
the mass of the black hole $M \sim T^4$, there
is no difference in scaling between the four- and five-dimensional 
pressures, and hence $3p/T^4$ in (\ref{fourdpre}) 
is initially approximately constant and then increases slightly
(due to the smallness of $\delta$) as the temperature increases.
Thus the pressure curve does not increase as 
abruptly as the energy density, and always lies beneath it
as long as it can be calculated reliably. 

As the temperature increases towards the upper end of the phase 
transition,
the increase in the pressure may be obtained from 
terms that have been ignored so far in deriving (\ref{eqstate}).
These include terms that express fluctuations of the potential 
${\overline U}$ with the volume $\Omega _4 - \Omega _0$.
These are required by continuity 
between the two asymptotic regimes for the pressure computed above.  
The generic (approximate) form of such terms may be found by representing 
the potential fluctuations as
\be
  U \simeq   {\overline U} + \epsilon '\frac{N}{\Omega _4 - \Omega _0}
\label{fluct}
\ee
where $\epsilon' $ is  small and positive.
Such a dependence of the potential on $\Omega _4 $ 
results in 
extra terms on the right-hand side of the equation of state.
Thus, for example, in the high-temperature phase 
we expect the the boundary pressure to have
the form:
\be
      p/T^4  \sim {\rm const} \frac{1}{\Delta \Omega _3 T^4 }\left[NT + 
\epsilon' N^2 (M + T)/(\Omega_4 - \Omega _0) \right]\;. 
\label{prs}
\ee
We now recall that, on the high-temperature 
side of the phase transition,  $\Delta \Omega _3$ 
scales like $T^{-3}$, $(\Omega _4 - \Omega _0)$ scales like $T^{-4}$,
and the mass of the black hole scales like $T^{-2}$.
The mass is sufficiently large that
the $M$-dependent term is still dominant. Hence,
from (\ref{fluct}) one obtains a linear increase for $3p/T^4$:
\be
   3p/T^4\sim{\rm const}+{\cal O}\left({\rm const'}\times\epsilon'N^2T
\right)\;.
\label{newpr}
\ee
As the temperature increases, the $\epsilon'$ term becomes smaller and 
smaller, and one recovers a constant temperature at the end of the
of the transition region.  
The proportionality constants are again fixed by 
requiring that this scaling should describe at
high temperature an almost ideal gas of massless
particles, in which case we have the relation $\rho = 3 p$
in three space dimensions.

We display in Fig.~\ref{threepre} heuristic
interpolations of $\rho$ and $3p$ between the high-
and low-temperature limits (i) and (ii).
These curves can be
compared with those calculated for QCD on the lattice~\cite{satz}.
In both cases,
there appears to be a sharp jump of the energy density at the 
onset of the deconfining phase transition,
from the value where the system is equivalent to a gas of pions, towards 
that where the system is described by an
almost ideal gas of quarks and gluons. 
On the other hand, the increase in the pressure is much smoother
in both the lattice and our AdS calculations.
This is related in our approach to the weak-field 
assumption for the potential: $|{\rm U}|\ll 1$, which
is valid for near-horizon AdS geometries in the high-temperature 
phase. 

We should repeat that our analysis in the limit (ii) is not yet
quantitative at low temperatures. However,
the Liouville world-sheet approach of~\cite{em}, which
describes the dynamics of world-sheet defects via
$D$ particles, describes 
this regime qualitatively correctly, leading in particular to
confinement as a low-temperature property. 
In this case, the space-time obtained from $D$-particle recoil is indeed
of AdS type with $M \rightarrow 0$ in (\ref{adsbh})~\cite{kanti,em}.  
We have shown in this paper that this approach has, moreover, a
plausible regular
continuation to the high-temperature limit (i) explored in~\cite{witt}.
This gives us further reason to hope that the Liouville string
approach may be suitable for development into a reliable tool for
describing non-perturbative gauge dynamics, and therefore may
contribute to the new avenue for non-perturbative gauge-theory
calculations opened up in~\cite{malda,witt,balls,potential,vacuum}.

\newpage

{\bf Acknowledgements}

One of us (A.G.) thanks the World Laboratory for a John Bell Scholarship.
Another (N.E.M.) thanks Mike Teper for useful discussions.
%, and the
%Swiss air-traffic controllers for making it possible to complete this
%paper in a timely manner.
The results of this paper were presented by J.E. at the memorial meeting
for Klaus Geiger: `RHIC Physics and Beyond - Kay-Kay-Gee Day',
BNL, October 23rd, 1998~\cite{KKG}. We dedicate this paper to
our friend Klaus's memory.

\end{document}